\documentclass[iop,numberedappendix]{emulateapj}
\slugcomment{Accepted by \apj} 
\newcommand{\di}{\mathrm{d}} 
\usepackage{natbib}
\usepackage{amsmath}
\usepackage{booktabs}
\usepackage{color}
\usepackage{graphicx, subfigure}
\usepackage{CJKutf8}
\defcitealias{WL2019}{WL2019}

\begin{document}
\title{Chondrule Formation by the Jovian Sweeping Secular Resonance}
\author{Munan Gong (\begin{CJK*}{UTF8}{gbsn}龚慕南\end{CJK*})\altaffilmark{1},
    Xiaochen Zheng (\begin{CJK*}{UTF8}{gbsn}郑晓晨\end{CJK*})\altaffilmark{2}, 
Douglas N.C. Lin (\begin{CJK*}{UTF8}{gbsn}林潮\end{CJK*})\altaffilmark{2,3,4},
Kedron Silsbee\altaffilmark{1}, 
Clement Baruteau\altaffilmark{5}, 
Shude Mao (\begin{CJK*}{UTF8}{gbsn}毛淑德\end{CJK*})\altaffilmark{2}}
\altaffiltext{1}{Max-Planck Institute for Extraterrestrial Physics,
Garching by Munich, 85748, Germany; 
munan@mpe.mpg.de}
\altaffiltext{2}{Department of Astronomy and Center for Astrophysics, 
Tsinghua University, Beijing 10086, China}
\altaffiltext{3}{Institute for Advanced Studies, 
Tsinghua University, Beijing 10086, China}
\altaffiltext{4}{Department of Astronomy and Astrophysics, University of
California Santa Cruz, Santa Cruz, CA 95064, USA}
\altaffiltext{5}{Institut de Recherche en Astrophysique
et Planétologie (IRAP) 14 avenue Edouard Belin, 31400 Toulouse, France}
\begin{abstract}
    Chondrules are silicate spheroids found in meteorites, serving as important
    fossil records of the early solar system. In order to form chondrules, 
    chondrule precursors must be heated to temperatures much higher than the typical
    conditions in the current asteroid belt.
    One proposed mechanism for chondrule heating is the passage through bow
    shocks of highly eccentric planetesimals in the protoplanetary
    disk in the early solar system. However, it is difficult for planetesimals to gain
    and maintain such high eccentricities.
    In this paper, we present a new scenario in which planetesimals in the
    asteroid belt region are excited to high eccentricities by the 
    Jovian sweeping secular resonance in a depleting disk,
    leading to efficient formation of chondrules.
    We study the orbital evolution of planetesimals in the
    disk using semi-analytic models and numerical simulations. We investigate
    the dependence of eccentricity excitation on the planetesimal's size as
    well as the physical environment, and calculate the probability for chondrule formation.
    We find that $50-2000~\mathrm{km}$ planetesimals can obtain eccentricities
    larger than $0.6$ and cause effective chondrule heating. Most chondrules form in
    high velocity shocks, in low density gas, and in the inner disk. The fraction
    of chondrule precursors which become chondrules 
    is about $4-9\%$ between $1.5-3~\mathrm{AU}$. 
    Our model implies that the disk depletion timescale is
    $\tau_\mathrm{dep}\approx 1~\mathrm{Myr}$, 
    comparable to the age spread of chondrules; and that Jupiter
    formed before chondrules, no more than $0.7~\mathrm{Myr}$ after the
    formation of the CAIs.
\end{abstract}

\section{Introduction}
Chondritic meteorites, or chondrites, contain some of the oldest and most primitive
solids in our solar system. They record the physical conditions at the early phase of
the solar nebula evolution and planet formation.
Chondrites are mainly composed of chondrules, which are 0.1 to 1
millimeter sized silicate spheroids. 
The majority of chondrules formed $1-3~\mathrm{Myr}$ after the calcium aluminum
inclusions (CAIs), another component often found in chondrites and 
the oldest solids dated in the solar system \citep{Scott2007}.
To form the chondrules we see today, chondrule
precursors need to be heated to temperatures above $1600~\mathrm{K}$
and cooled down rapidly on timescales of minutes to hours \citep{Desch2012}.
Considering ordinary chondrites make up $\sim 10\%$ of meteorites found on
Earth \citep{Desch2005}, chondrule heating events should be common in 
the early solar system.  

Many mechanisms have been proposed for chondrule heating, including solar flares
\citep{Shu2001}, impacts between planetesimals \citep{UC1953, Dullemond2014},
and shock heating in the protoplanetary disk around the Sun
\citep{Iida2001, CH2002, Desch2005}. 
Chondrule heating in
planetesimals' bow shocks is one of the most promising models that can
simultaneously explain many features in the thermal histories of chondrules,
such as the ambient temperature, the peak temperature, and the cooling rate
\citep{Desch2012}.
In recent years, detailed numerical simulations that model the structure of the
bow shocks and the trajectories of chondrules confirmed that
passing through the bow shocks of large planetesimals is a
plausible mechanism for chondrule formation \citep{Morris2012, Mann2016}.
However, there is one remaining puzzle in this model: in order to heat up chondrule
precursors to the required high temperatures, the relative velocity between the
planetesimal and the gas needs to be 
$v_\mathrm{rel} \gtrsim 6~\mathrm{km/s}$ in
relatively dense gas, and even higher if the gas density is lower
\citep{Iida2001, Mann2016}. This is a significant fraction ($24-37\%$) 
of the Keplerian
speed in the asteroid belt region, and implies that the planetesimals
responsible for chondrule heating need to be excited to high eccentricities.

One event that can lead to the eccentricity excitation of planetesimals is the
formation of Jupiter. As the largest planet in the solar system, Jupiter can
profoundly affect the dynamics of the planetesimals. With a thick
gaseous atmosphere, Jupiter should have formed early, before the gas in the
disk was significantly depleted.  
\citet{Weidenschilling1998} proposed that the strong 2:1 and 3:2 
mean motion resonances with
Jupiter can excite the eccentricities of  planetesimals to $e_p \approx 0.3$.
However, because the resonance locations are relatively far out in the disk 
(at $3.28$ and $3.97~\mathrm{AU}$) where the Keplerian velocity is low
($\lesssim 16~\mathrm{km/s}$), the planetesimals can only obtain velocities of
$v_\mathrm{rel} \lesssim 5~\mathrm{km/s}$, not high enough for chondrule formation.
\citet{Nagasawa2014} considered in addition the gravity of the disk which
enables the secular resonance by Jupiter. They found that planetesimals at the
location of resonance
can gain $v_\mathrm{rel} > 12~\mathrm{km/s}$ and lead to effective chondrule
formation. However, they only considered a fixed disk mass and thus a fixed
location of the Jovian secular resonance, and only planetesimals of a
single size of $300~\mathrm{km}$. 

In this paper, we investigate a scenario in which the orbits of planetesimals are
influenced by both Jupiter and a depleting protoplanetary disk. As the disk mass
decreases with time, the location of the Jovian secular resonance moves from
the outer to the inner disk. This phenomenon is
called the sweeping secular resonance, first proposed by \citet{Ward1976} to
explain Mercury's large eccentricity and inclination.\footnote{The sweeping
secular resonance discussed in \citet{Ward1976} is caused by the decreasing
oblateness of the Sun as its spin slows down, rather than the disk depletion as
studied in this paper.}
Subsequently, the
sweeping secular resonance was invoked to explain other characteristics of the solar
system, such as the formation and orbits 
of terrestrial planets \citep{NLT2005, TNL2008},
the mass deficit of the asteroid belt region in the minimum mass solar nebula 
(MMSN) model \citep{Hayashi1981, Zheng2017}, and the size distribution of the
asteroids \citep{Zheng2017}.
We propose that the same mechanism can explain 
chondrule formation, by exciting the planetesimals in the asteroid belt
region to high eccentricity orbits. We use both semi-analytic models and
numerical simulations to investigate the 
dependence of eccentricity excitation on the planetesimal's size, as well as
the environment and probability of chondrule formation.

The structure of this paper is as follows. Section \ref{section:methods}
describes the method of our semi-analytic models and numerical simulations.
In Section \ref{section:results}, we show our results on the orbital evolution of
planetesimals, the size-dependent eccentricity excitation,
and the chondrule formation probability and its dependence on the model parameters.
In Section \ref{section:discussions},
we discuss the implications of chondrule formation on the formation of Jupiter
and the depletion timescale of the disk. Finally, Section \ref{section:summary}
summarises the main findings of this work.

\section{Methods}\label{section:methods}
\subsection{Semi-analytic Model \label{section:analytic}}
In this section, we describe our semi-analytic model for 
the eccentricity and semi-major axis evolution of planetesimals that are
embedded in the protoplanetary disk and perturbed 
by the Jovian sweeping secular resonance.
The secular perturbation by Jupiter excites the eccentricity of
planetesimals, while gas drag damps both their 
eccentricities and semi-major axes. We assume the Sun, disk, planetesimals, and
Jupiter are all in the same plane, and thus our model is 2-dimensional.

We consider the stage when Jupiter has already gained its present-day mass and
opened a gap in the disk which quenched its accretion \citep{DobbsDixon2007}.
In this model, we assume that: 
(1) There is a thin, axisymmetric protoplanetary disk with power-law surface density
distribution based on the MMSN model, with a gap near Jupiter's orbit
\citep{Bryden1999}. The surface density of the disk depletes over time.
(2) The masses of planetesimals are much smaller than the mass of Jupiter, and
therefore, the gravitational perturbation by planetesimals on Jupiter is
ignored. In our fiducial model, we do not consider the presence of Saturn, and 
Jupiter's semi-major axis and eccentricity are at their current day
values ($a_J=5.2~\mathrm{AU}$, $e_J=0.05$), and do not evolve with time.
The only time variation in Jupiter's orbit is its apsidal precession due to the
gravity of the gas disk.  
(3) All planetesimals are initially embedded in the disk with 
circular orbits in the region of the present-day asteroid belt. 
Their subsequent orbital evolution is affected
by the gravitational perturbation from Jupiter and the surrounding gas disk,
as well as the gas drag from the disk.

Later, we introduce Saturn in some of our numerical simulations. Although Saturn
introduces its own additional secular resonance and perturbs the orbit of
Jupiter, we show that the excitation of planetesimals' eccentricities is still
dominated by the Jovian sweeping secular resonance,
and the chondrule formation probability is largely unchanged.

\begin{figure*}[htbp]
\centering
\includegraphics[width=\linewidth]{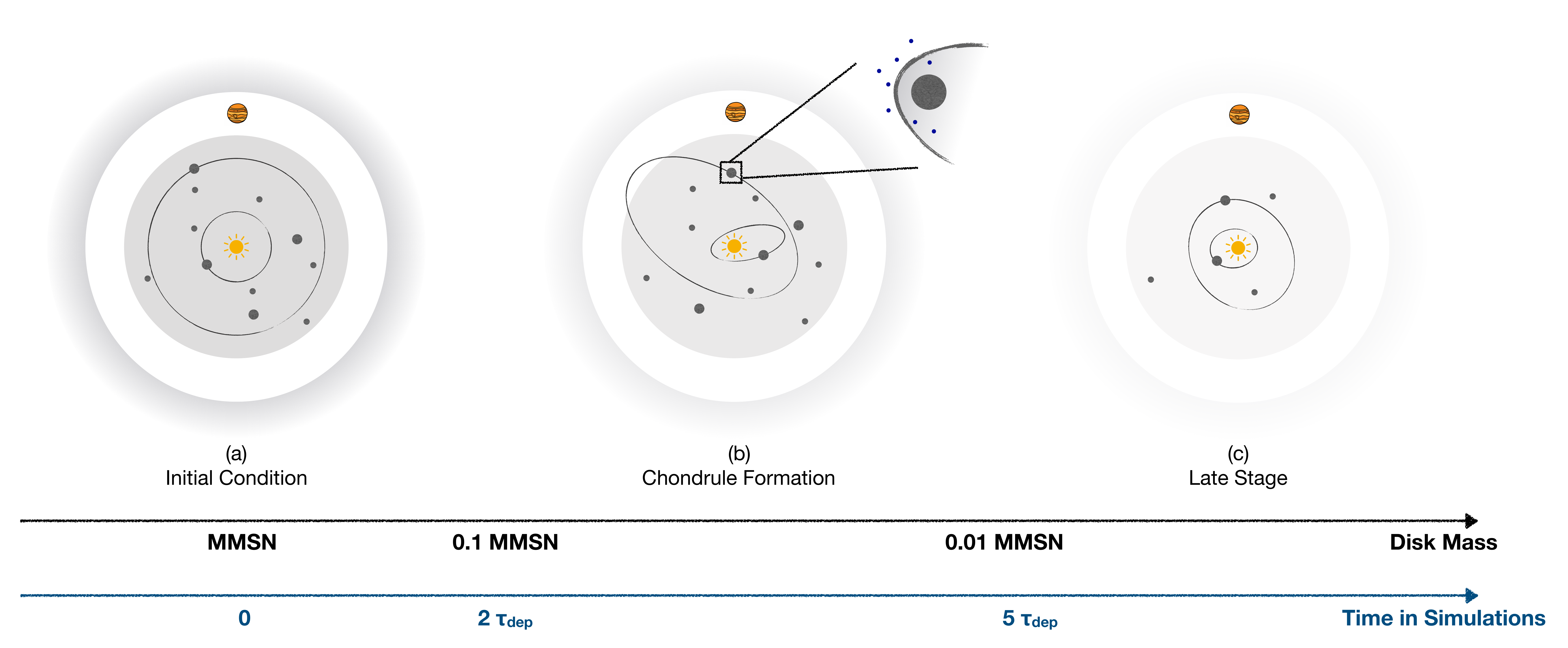}
    \caption{Schematic picture of our chondrule formation model. (a) Initial
    condition: Jupiter forms and opens a gap in the disk. The planetesimals are in
    circular orbits. (b) Chondrule formation: As the gas disk depletes over time, 
    the sweeping secular resonance by Jupiter excites the eccentricities of
    planetesimals. Chondrules form in the bow shocks of highly eccentric
    planetesimals. (c) Late stage: The gas disk is almost fully depleted. Many
    planetesimals are ejected from the system by close encounters with Jupiter
    or removed from our simulation when they pass too close to the Sun.
    The remaining planetesimals are in orbits with moderate
    eccentricities.}
\label{fig:story}
\end{figure*}

The physical picture of our model (without Saturn)
is illustrated in Figure \ref{fig:story}.
We focus on the process of chondrule formation in this paper.
\citet{Zheng2017} adopted a similar physical picture, but focused on the
remaining distribution of planetesimals in the late stage after the gas disk is
almost fully depleted.

Therefore, the equations for orbital evolution of Jupiter and the planetesimals can be
written as:
\begin{align}
    &\frac{\di e_p}{\di t} = 
       \frac{\di e_{p, J}}{\di t} + \frac{\di e_{p, \mathrm{gas}}}{\di
       t}\label{eq:ode1}\\
    &\frac{\di \xi}{\di t} \equiv \frac{\di (\varpi_p - \varpi_J)}{\di t} = 
                       \frac{\di \varpi_{p,J}}{\di t} 
                       + \frac{\di \varpi_{p,\mathrm{disk}}}{\di t}
                       - \frac{\di \varpi_{J, \mathrm{disk}}}{\di t}
                       \label{eq:detadt} \\
    &\frac{\di a_p}{\di t} = \frac{\di a_{p, \mathrm{gas}}}{\di t},
                       \label{eq:ode3}
\end{align}
where $e_p$, $a_p$ and $\varpi_p$ ($e_J$, $a_J$ and $\varpi_J$) are the eccentricity,
semi-major axis, and apsidal angle of the planetesimals (Jupiter). 
Given that the combined mass of 
planetesimals is much less than that of Jupiter, we can ignore the
gravitational perturbation of planetesimals on Jupiter. In the sections below,
we describe in detail the different terms in
Equations (\ref{eq:ode1}), (\ref{eq:detadt}), and (\ref{eq:ode3}). These
equations can be solved numerically as a set of coupled ordinary differential
equations.

\subsubsection{Sweeping Secular Resonance by Jupiter}
The eccentricity vector of a planetesimal is modulated by the gravitational
force from Jupiter. When $e_{p, J} \ll 1$,
to the first order \citep{MD1999,NLI2003}:
\begin{align}
    &\frac{\di e_{p, J}}{\di t} = \frac{e_J}{t_p}C\sin\xi, \label{eq:depJdt}\\
    &\frac{\di \varpi_{p,J}}{\di t} = 
    \frac{1}{t_p}\left[ \frac{e_J}{e_p}C\cos\xi +1 \right].
\end{align}
Here $C=-b_{3/2}^{(2)}(\alpha)/b_{3/2}^{(1)}(\alpha) $ and
$t_p = [4/(n_p b_{3/2}^{(1)}(\alpha) \alpha^2)] (M_\sun/M_J)$,
where $n_p = \sqrt{GM_\sun/a_p^3}$ is the mean motion of
the planetesimals, $\alpha \equiv a_p/a_J$ is the ratio of semi-major axes
between the planetesimal and Jupiter, and
$b_s^{(j)}$ are the Laplace coefficients. 

The orbits of planetesimals and Jupiter precess due to the 
gravitational potential of the disk. We consider a thin disk with
a power law surface density profile,
\begin{equation}\label{eq:surf_dens}
    \Sigma (R) = f_g \Sigma_0 \left( \frac{R}{1\mathrm{AU}} \right)^{-k},
\end{equation}
where $R$ is the radial coordinate. 
The initial disk surface density is set to match the MMSN model with
$\Sigma_0=1700~\mathrm{g~cm}^{-2}$ and $k=3/2$ \citep{Hayashi1981}.
In order to mimic the depletion of the disk over time, we set the disk surface
density to decay exponentially over timescale $\tau_\mathrm{dep}$:
\begin{equation}\label{eq:f_g}
    f_g =  \exp \left(-\frac{t}{\tau_\mathrm{dep}}\right).
\end{equation}
Observations of young clusters suggest that most stars lose their disk on a 
timescale of $\leq 3~\mathrm{Myr}$ \citep{Zuckerman1995, Haisch2001}. 
This idealized prescription for disk evolution does not take into account 
the possibility of interruption of gas accretion flow in the disk due to
planets' tidal torque or photoevaporation at some critical radii.  
Nonetheless, it is useful to illustrate the consequence of gas depletion in the
disk.
We set the parameter $\tau_\mathrm{dep}=1~\mathrm{Myr}$ in our fiducial model, 
and explore the effect of varying $\tau_\mathrm{dep}$ 
in Section \ref{section:probability}.

The precession of Jupiter due to the disk with a power-law surface density
profile (Equation (\ref{eq:surf_dens})) and a gap surrounding Jupiter
is given by \citep{Ward1981}:
\begin{align}
    \frac{\di \varpi_{J, \mathrm{disk}}}{\di t} 
    &= \frac{2\pi G \Sigma(a_J)}{n_J a_J} \sum_{n=1}^{\infty} n(2n+1)A_n \label{eq:dw_J,disk}\\
    & \times \left[  \frac{(a_J/R_\mathrm{out})^{2n-1+k}}{2n-1+k}   
      + \frac{(R_\mathrm{in}/a_J)^{2n+2-k}}{2n+2-k} 
  \right] \nonumber\\
  &= \left.\frac{\di \varpi_{J, \mathrm{disk}}}{\di t}\right\vert_{t=0} 
  \exp \left(-\frac{t}{\tau_\mathrm{dep}}\right),\nonumber
\end{align}
and the planetesimal precession rate due to the disk is given by
\begin{align}
    \frac{\di \varpi_{p, \mathrm{disk}}}{\di t} 
    &= \frac{2\pi G \Sigma(a_p)}{n_p a_p} \left(
        -\frac{1}{2} + \frac{S_k}{2} + S_{\mathrm{out}} - S_{\mathrm{in}}
    \right)\label{eq:dw_p,disk}\\
    &=\left.\frac{\di \varpi_{p, \mathrm{disk}}}{\di t}\right\vert_{t=0}
      \exp \left(-\frac{t}{\tau_\mathrm{dep}}\right).\nonumber
\end{align}
Here $R_\mathrm{in}$ and $R_\mathrm{out}$ are the radii of the
inner and outer edges of the
gap, and $A_n = [(2n)!/(2^{n}n!)^2]^2$. 
We choose  $R_\mathrm{in} = 4.5~\mathrm{AU}$ and $R_\mathrm{out} =
6~\mathrm{AU}$ based on numerical simulations by \citet{Bryden1999}.\footnote{
    We use the results from their model 1B with the disk scale-height
    $H/R=0.04$ and the turbulent viscosity parameter $\alpha=0.001$
    \citep{SS1973}. More recent simulations by \citet{DK2015} also obtained
    similar results.}
In Equation (\ref{eq:dw_p,disk}), the constant term,
\begin{equation}
    S_k = (k-1)(k-2)\sum_{n=1}^{\infty} A_n \frac{4n+1}{(2n-1+k)(2n+2-k)}, 
\end{equation}
and $S_k = -0.094$ for $k=3/2$.
The other two terms, $S_{\mathrm{out}}$ and $S_{\mathrm{in}}$, are from the
inner and outer edges of the gap,
\begin{align}
    S_{\mathrm{out}} &= \sum_{n=1}^{\infty}n(2n+1)A_n
    \frac{(a_p/R_\mathrm{out})^{2n-1+k}}{2n-1+k},\\
    S_{\mathrm{in}} &= \sum_{n=1}^{\infty}n(2n+1)A_n
    \frac{(a_p/R_\mathrm{in})^{2n-1+k}}{2n-1+k}.
\end{align}

\begin{figure}[htbp]
\centering
\includegraphics[width=\linewidth]{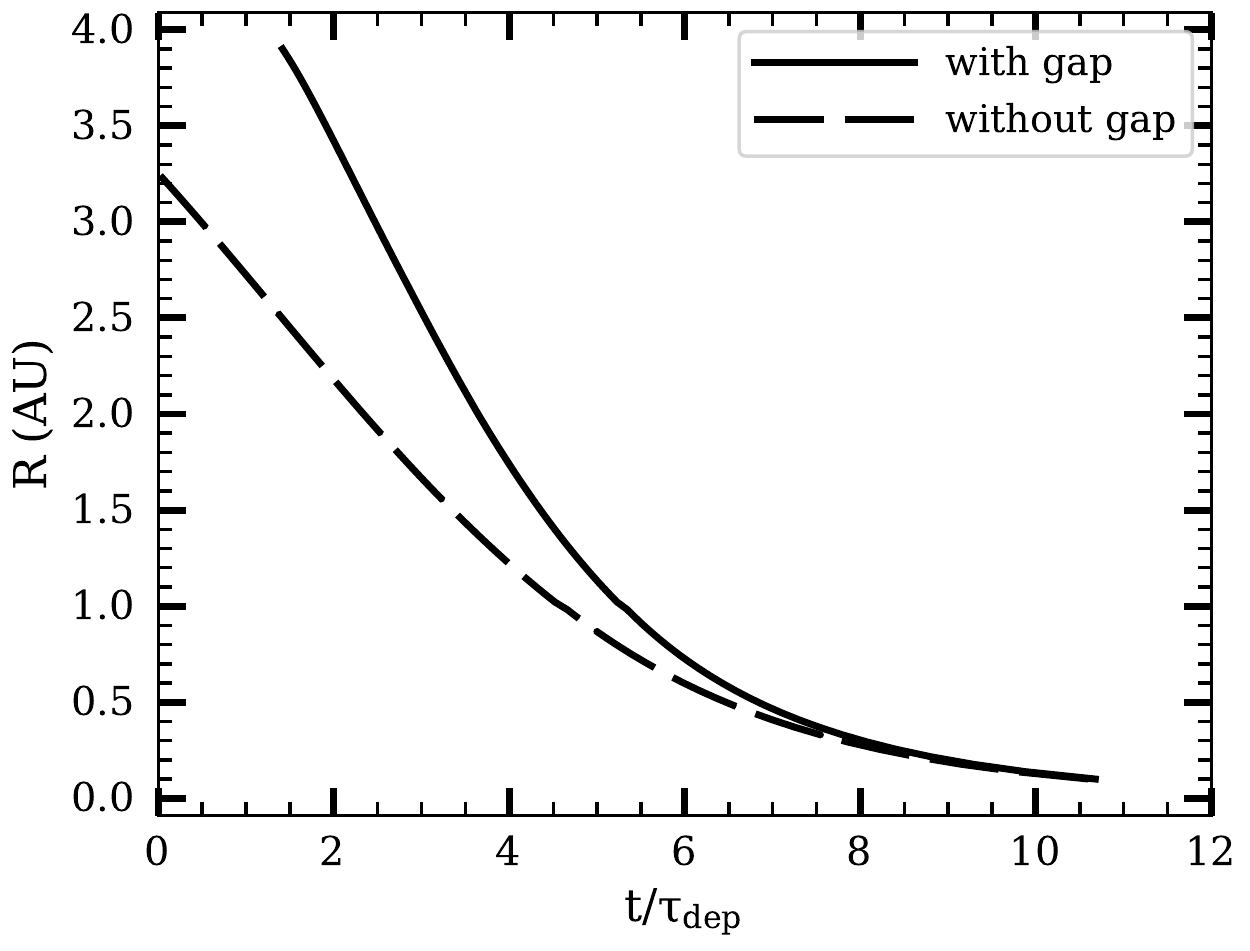}
\caption{Location of Jupiter's secular resonance $\nu_5$ with a gap in the disk
    between $4.5-6~\mathrm{AU}$ (solid) and without a gap (dashed).}
\label{fig:nu5}
\end{figure}

The secular resonance between a planetesimal and Jupiter occurs when 
their apsidal precession rates coincide, leading to the growth of the
planetesimal's eccentricity. The location of Jovian resonance $\nu_5$ can be
solved analytically:
setting $\di \xi /\di t = 0$ in Equation (\ref{eq:detadt}) and combining with
Equations (\ref{eq:dw_J,disk}) and (\ref{eq:dw_p,disk}), the time $t$ when
resonance occur at a radius $R$ is given by
\begin{equation}
    \frac{t}{\tau_\mathrm{dep}} = -\ln \left( \frac{
      \frac{\di \varpi_{p, \mathrm{J}}}{\di t}}{
          \frac{\di \varpi_{J, \mathrm{disk}}}{\di t}|_{t=0}
  - \frac{\di \varpi_{p, \mathrm{disk}}}{\di t}|_{t=0}} \right),
\end{equation}
where the numerator is evaluated at time $t=0$. The location of $\nu_5$ as a
function of time is shown in Figure \ref{fig:nu5}.  
As the disk surface density depletes over time, $\nu_5$ 
moves closer to the Sun. The gap carved by Jupiter delays
the time when $\nu_5$ passes a particular location:
at $t\lesssim\tau_\mathrm{dep}$, there is no
resonance inside $\sim 4~\mathrm{AU}$. Jupiter's secular resonance only 
starts to stir up the planetesimals after the disk is significantly depleted. 

\subsubsection{Damping by the Gas Disk\label{section:gas-damping}}
The gas in the disk damps the eccentricity and semi-major axis of a
planetesimal through aerodynamic and tidal effects.
We follow \citet{ZL2007} to estimate the gas drag. The formulae only
include the lowest order terms assuming $e_p, \eta \ll 1$ 
($\eta$ is a parameter related to the pressure gradient,
see Equation (\ref{eq:eta})). 

The acceleration of a planetesimal with mass $m_p$ and radius $r_p$ by
aerodynamic drag is
\begin{equation}\label{eq:f_aero}
    \mathbf{f}_\mathrm{aero} = - \frac{1}{2m_p}C_D \rho_g \pi r_p^2
    |\mathbf{v}_\mathrm{rel}|\mathbf{v}_\mathrm{rel},
\end{equation}
where $C_D = 0.44$ is the coefficient for objects with large
Reynolds number \citep{Whipple1972}, $\rho_g$ is the gas density, 
and $\mathbf{v}_\mathrm{rel} =
\mathbf{v_k} - \mathbf{v_g}$ is the relative velocity between the Keplerian
motion of the planetesimal ($\mathbf{v_k}$) and the sub-Keplerian velocity of the gas
($\mathbf{v_g}$) due to pressure gradients in the disk. Because 
planetesimals are expected to settle near the disk mid-plane, we
use the gas density at the mid-plane for calculating the aerodynamic drag. The disk
vertical ($\hat{z}$) structure is set by the balance between gas pressure and
gravity, giving
\begin{equation}\label{eq:rho_g_z}
    \rho_g(R, z) = \rho_g(R,0)\exp\left( - \frac{z^2}{2H^2} \right),
\end{equation}
where the disk scale height $H = c_s / \Omega_k$, 
$c_s = \sqrt{k_B T/\mu m_H}$ is the
sound speed and $\Omega_k = \sqrt{GM_\sun/R^3}$ is the Keplerian angular
velocity, $m_H$ is the proton mass, and $k_B$ is the Boltzmann constant.
The mean molecular weight of the gas $\mu=7/3$, assuming the disk to be made up
of $\mathrm{H_2}$ and $\mathrm{He}$ with the solar abundance
$n(\mathrm{H_2}):n(\mathrm{He})=5:1$. 
We adopt the temperature profile of the disk in the MMSN model
\citep{Hayashi1981}:
\begin{equation}
    T = T_0 \left( \frac{R}{1\mathrm{AU}}\right)^{-1/2}
    = 280 ~\mathrm{K}\left( \frac{R}{1\mathrm{AU}}\right)^{-1/2},
\end{equation}
which gives the disk scale-height
\begin{equation}
    \frac{H}{R} = 0.033 \left( \frac{R}{1\mathrm{AU}} \right)^{1/4}
                       \left( \frac{T_0}{280\mathrm{K}} \right)^{1/2},
\end{equation}
and gas density at the mid-plane
\begin{align}\label{eq:rho_g}
    \rho_g(R, 0) &= \frac{\Sigma}{\sqrt{2\pi} H} 
    = 1.36\times 10^{-9}~\mathrm{g~cm}^{-3}\\
    &\times f_g \left(\frac{R}{1\mathrm{AU}}\right)^{-11/4} 
    \left(\frac{T_0}{280\mathrm{K}}\right)^{-1/2}.\nonumber
\end{align}

The effect of aerodynamic gas drag on the secular evolution of a planetesimal's orbit
is derived by \citet{Adachi1976} (to the lowest order in $e_p$ and $\eta$), 
assuming the gas in the disk follows circular orbits: 
\begin{align}
\frac{1}{a_p}\left( \frac{\di a_p}{\di t} \right)_\mathrm{aero}
&= -\frac{2}{\tau_\mathrm{aero}} \left( \frac{5}{8}e_p^2 + \eta^2 \right)^{1/2}
\left(  e_p^2 + \eta \right) \label{eq:dapdt_aero}\\
\frac{1}{e_p}\left( \frac{\di e_p}{\di t} \right)_\mathrm{aero}
&= -\frac{1}{\tau_\mathrm{aero}} \left( \frac{5}{8}e_p^2 + \eta^2 \right)^{1/2},
\label{eq:depdt_aero}
\end{align}
where $\tau_\mathrm{aero}$ is the aerodynamic drag timescale given by
\begin{align}\label{eq:tau_aero}
    &\tau_\mathrm{aero} = \frac{2m_p}{\pi C_D r_p^2 \rho_g v_k}\\
                       &= \frac{4.75}{f_g} \mathrm{yr}
    \left(\frac{R}{1\mathrm{AU}}\right)^{13/4}
    \left(\frac{r_p}{1\mathrm{km}}\right) 
    \left(\frac{\rho_p}{1\mathrm{g/cm}^3}\right)
    \left(\frac{T_0}{280\mathrm{K}}\right)^{1/2},
\end{align}
$\eta$ is a parameter related to the radial pressure
gradient in the disk
\begin{equation}\label{eq:eta}
    \eta(R) = \frac{13}{8}\left(\frac{H}{R}\right)^2 
    = 1.5\times 10^{-3} \left( \frac{R}{1\mathrm{AU}}\right)^{1/2}
    \left( \frac{T_0}{280\mathrm{K}}\right),
\end{equation}
and $v_k$ the Keplerian velocity of a circular orbit $v_k =
\sqrt{GM_\sun/R}$.

The densities of planetesimals are expected to be in the range of $\sim
1-5.5\mathrm{g~cm^{-3}}$. Small planetesimals may be considered as 
pebble/ice piles, while large planetesimals may have gone through
differentiation processes, similar to iron/stone meteorites. 
We use a simple
prescription for the densities of planetesimals, following \citet{Zheng2017}: 
\begin{equation}
    \rho_p(r_p)=\begin{cases}
        1.0~\mathrm{g~cm^{-3}}, & \text{$r_p \leq 18\mathrm{km}$}, \\
        \frac{r_p}{18\mathrm{km}}~\mathrm{g~cm^{-3}}, &
        \text{$18\mathrm{km} < r_p < 100\mathrm{km}$}, \\
        50/9~\mathrm{g~cm^{-3}}, & \text{$r_p \geq 100\mathrm{km}$}.
  \end{cases}
\end{equation}

In addition to the aerodynamic drag, large planetesimals also experience 
the tidal drag force by the disk through Lindblad resonances
\citep{Ward1988, Artymowicz1993, TNL2008}. The typical  
timescale for tidal damping by Lindblad torque is given by
\begin{align}\label{eq:tau_tidal}
    \tau_\mathrm{tidal} &= \left( \frac{m_p}{M_\sun} \right)^{-1}
    \left( \frac{\Sigma R^2}{M_\sun} \right)^{-1}
    \left( \frac{c_s}{v_k} \right)^4 \Omega_k^{-1}\\
    &=\frac{4.92\times 10^{5}~\mathrm{yr}}{f_g}  \nonumber\\
    &\times \left( \frac{\rho_p}{1\mathrm{g/cm}^3} \right)^{-1}
    \left( \frac{r_p}{10^3\mathrm{km}} \right)^{-3}
    \left( \frac{R}{1\mathrm{AU}} \right)^2
    \left( \frac{T_0}{280\mathrm{K}} \right)^2. \nonumber
\end{align}
In addition to the Lindblad resonances, the corotation torques
also contribute to the drag force \citep{Paardekooper2011}. 
With the disk profile adopted in this work, the corotation torque 
is small compared to the Lindblad torque for
circular orbits \citep{Paardekooper2011}. Furthermore, numerical simulations in 
\citet{FN2014} show that corotation torque decreases with increasing
eccentricity. Therefore, we neglect the corotation torque in our models.

Numerical simulations of the orbital evolution of planetesimals in a gaseous disk  
obtained a tidal damping timescale similar to the
analytic expression in Equation (\ref{eq:tau_tidal}) for planetesimals with low
eccentricities \citep{PL2000}, and different tidal damping timescales depending
on the disk structure for highly-eccentric planetesimals \citep{Muto2011}.
However, the results from these 2-dimensional numerical simulations 
depend sensitively on the numerical softening parameter, which is uncertain and
arbitrary. Therefore, we choose to use the simple analytic expression in 
Equation (\ref{eq:tau_tidal}). To investigate the dependence of our results on the
tidal damping, we run an additional
numerical simulation (TDL5 in Table \ref{table:models})
with five times stronger tidal damping (multiplying $\tau_\mathrm{tidal}$ in
Equation (\ref{eq:tau_tidal}) by a factor of 0.2).

The acceleration of a planetesimal due to the tidal drag force can be estimated
using the formulation in \citet{KI2002},
\begin{equation}\label{eq:f_tidal}
    \mathbf{f}_\mathrm{tidal} =
    -\frac{\mathbf{v}_\mathrm{rel}}{\tau_\mathrm{tidal}}.
\end{equation}
Similar to the aerodynamic drag, the secular evolution of a planetesimal's
orbital elements due to the tidal drag is given by 
\citet{Adachi1976} (their Equation (4.18)):\footnote{In the limit of circular
orbits, Equation (\ref{eq:dapdt_tidal}) does not recover the typical formulae
for type I migration \citep[e.g.][]{Paardekooper2011,Baruteau2014}. However, 
here we focus on the semi-major axis and eccentricity damping of planetesimals
with high eccentricities.}
\begin{align}
\frac{1}{a_p}\left( \frac{\di a_p}{\di t} \right)_\mathrm{tidal}
&= -\frac{2}{\tau_\mathrm{tidal}} \left( \eta + \frac{13}{16} e_p^2
\label{eq:dapdt_tidal} \right),\\
\frac{1}{e_p}\left( \frac{\di e_p}{\di t} \right)_\mathrm{tidal}
&= -\frac{1}{\tau_\mathrm{tidal}}.
\label{eq:depdt_tidal}
\end{align}

\begin{figure*}[htbp]
\centering
\includegraphics[width=\linewidth]{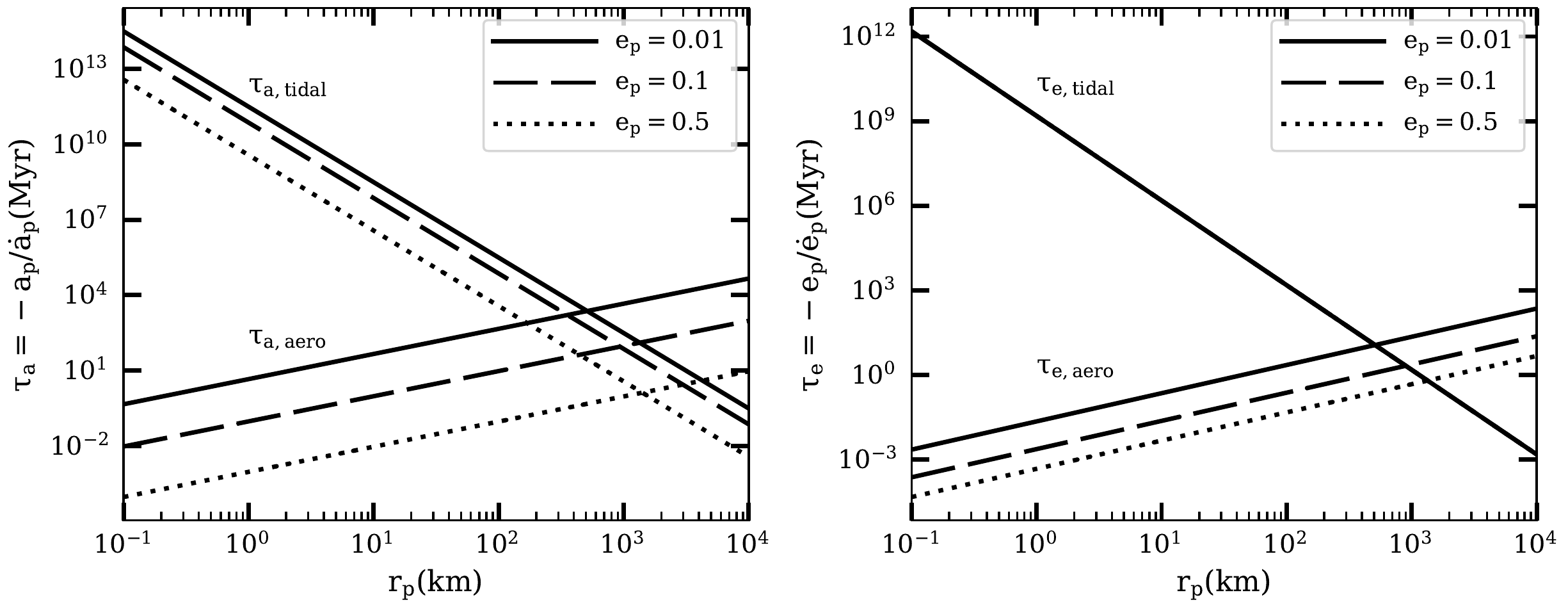}
\caption{Timescales
    for semi-major axis decay $\tau_a = -a_p/\dot{a}_p$ (left panel) and
    eccentricity decay $\tau_e = -e_p/\dot{e}_p$ (right panel) of planetesimals
    with radius $r_p=0.1-10^4~\mathrm{km}$ by aerodynamic or tidal gas drag.
    The planetesimals have semi-major axes $a_p=2.5\mathrm{AU}$.
    The solid, dashed and dotted lines show the cases in which the eccentricities of
    the planetesimals are $e_p=0.01, 0.1$ and $0.5$.}
\label{fig:tau}
\end{figure*}

Figure \ref{fig:tau} shows the aerodynamic and tidal
damping timescales for planetesimals with $a_p=2.5\mathrm{AU}$ and sizes ranging 
from $0.1$ to $10^4$ km, as described by Equations 
(\ref{eq:dapdt_aero}), (\ref{eq:depdt_aero}), (\ref{eq:dapdt_tidal}) and
(\ref{eq:depdt_tidal}). When the size of a planetesimal increases, 
the aerodynamic timescale increases whereas the tidal timescale decreases.
We can roughly divide the planetesimals into three groups according to the
gas-damping that they experience: planetesimals with $r_p \lesssim
50~\mathrm{km}$ are subject to ``strong aerodynamic damping'', $r_p \sim
50-2000~\mathrm{km}$ are ``weakly coupled'' to the gas, and $r_p \gtrsim
2000~\mathrm{km}$ are subject to ``strong tidal damping''. Planetesimals in
these different groups have very different behavior, which we discuss in detail
in Section \ref{sec:result:orbit}.
The planetesimals in the ``weakly coupled'' group are less susceptible 
to the damping of their eccentricity 
by gas drag, and can easily obtain high eccentricities by gravitational
interaction with Jupiter. 

\subsection{Numerical Simulation}
The analytic model in Section \ref{section:analytic} is based on two key
assumptions. First, Jupiter only affects the orbits of planetesimals by
secular gravitational perturbations. Other gravitational effects by Jupiter,
such as the mean motion resonance, are ignored.
Second, the formulations for the orbital evolution of planetesimals are
obtained by using linear perturbation theory, which breaks down when the
eccentricities of planetesimals are of order unity. 

To test these assumptions and to obtain a more complete picture 
of the orbital evolution of planetesimals, we carry out numerical simulations
using a modified version of the publicly available 
N-body code {\sl HERMIT4} \citep{Aarseth2003}. The simulation setup 
in our fiducial model is very
similar to the model $A_5$ in \citet{Zheng2017}. We briefly summarize the
method of our simulations here, and refer the readers to \citet{Zheng2017}
for more details.

The N-body code computes the gravitational interactions
between the Sun and Jupiter, and the gravitational forces from the Sun and
Jupiter to the planetesimals. We ignore the gravitational interaction from the
planetesimals back to the Sun and Jupiter, and between the planetesimals
themselves. Jupiter and the planetesimals also feel the gravitational potential
from the disk, which leads to precession of their orbits. The gas drag is
calculated by adding drag forces to the planetesimals' equations of motion,
as described in Equations (\ref{eq:f_aero}) and (\ref{eq:f_tidal}).

We place the planetesimals initially in circular orbits with random phase
angle between $0$ and $2\pi$. We put $2000$ planetesimals with semi-major
axes drawn from a random distribution between 1.5 and 3.5 AU.
The sizes of the planetesimals are randomly drawn
from a flat distribution in logarithm space between $10$ and $5000~\mathrm{km}$.
Later, when we calculate the chondrule formation probability, we scale the
number of planetesimals in each size bin to fit the realistic
size distributions of planetesimals (see Section \ref{section:probability}). 

We ran the simulation for 10 times the disk depletion time.
Because we are interested in
planetesimals that pass the main belt region of $R=1.5-3.5~\mathrm{AU}$, we
stop tracking a planetesimal once its semi-major axis is smaller than
$0.7~\mathrm{AU}$ or larger than $7~\mathrm{AU}$. 
For numerical reasons, we also remove planetesimals that
come within $0.07~\mathrm{AU}$ of the Sun.

In order to investigate the effects of our assumptions and parameters on
chondrule formation, we run a series of numerical models with different
setups, as summarized in Table \ref{table:models}. We explore different values
of the disk depletion time, tidal damping strength, and the initial
eccentricity of Jupiter. We also add Saturn in two models, with its semi-major
axis at its current day value or at $3:2$ mean motion resonance with Jupiter.
In models with Saturn, the gap is larger, encompassing the orbits of both
Jupiter and Saturn. Saturn interacts gravitationally with
the Sun, planetesimals, and the gas disk in the same way as Jupiter. The
gravitational interaction between Saturn and Jupiter is also included. 
The initial semi-major axis of Jupiter is always at its current day
value ($a_J=5.2~\mathrm{AU}$).

\begin{table}[htbp]
    \caption{Parameters for the Numerical Simulations}
    \label{table:models}
    \centering
    \begin{tabular}{ccc cc}
        \tableline
        \tableline
        model ID &$\tau_\mathrm{dep}~(\mathrm{Myr})$ 
        &$e_\mathrm{J,0}$\tablenotemark{a} 
        &gap$~(\mathrm{AU})$\tablenotemark{b} 
        &$a_{S,0}~(\mathrm{AU})$\tablenotemark{c}\\
        \tableline
        fiducial&1   &0.05 &4.5--6   &-\\
        DEP0p5  &0.5 &0.05 &4.5--6   &-\\
        DEP2    &2   &0.05 &4.5--6   &-\\
        TDL5\tablenotemark{d}     &1   &0.05 &4.5--6 &-\\
        EJ0p1    &1   &0.1  &4.5--6  &-\\
        ST       &1   &0.05 &4.5--11 &9.58\\
        STR      &1   &0.05 &4.5--11 &6.81\\
        \tableline
        \tableline
    \tablenotetext{1}{The initial eccentricity of Jupiter.}
    \tablenotetext{2}{The inner and outer radii
        of the gap opened by Jupiter (and Saturn,
        if it is in the model) in the gaseous disk.}
    \tablenotetext{3}{The initial semi-major axis of Saturn. ``-'' denotes that
        Saturn is not included in the model. Saturn's initial semi-major axis
        is at its current day value in model ST. In model STR, Saturn and
        Jupiter are in 3:2 mean-motion resonance, and we reduce Saturn's semi-major
        axis accordingly.}
    \tablenotetext{4}{In model TDL5, we increase the strength of tidal damping
        by a factor of 5 (multiplying $\tau_\mathrm{tidal}$ in 
        Equation (\ref{eq:tau_tidal}) by a factor of 0.2). 
        All other parameters are the same as the fiducial model.}
    \end{tabular}
\end{table}

\subsection{Probability of Chondrule Formation}\label{section:probability}
In order for chondrules to form in the bow shock of a planetesimal, the
relative velocity $v_\mathrm{rel}$ 
between the planetesimal and the gas has to be in a certain
range: the velocity needs to be high enough to melt chondrule precursors and low
enough to avoid complete evaporation. 
We use the result from \citet{Iida2001} (their Equations (37) and (38)) for the
velocity range required for chondrule formation.
This velocity range
depends on gas density, which affects the heating rate of chondrule precursors.

We use the following method to estimate the 
probability of chondrule formation.
For each planetesimal, the mass of chondrule precursors
going through the shock region per unit time is
\begin{equation}\label{eq:M_c}
    \frac{\di M_{c,\mathrm{heating}}}{\di t} = \sigma_c v_\mathrm{rel} n_c m_c,
\end{equation}
where $\sigma_c$ is the cross-section of the shock region, $v_\mathrm{rel}$
is the relative velocity between the planetesimal and the chondrule precursors,
$n_c$ is the number density of chondrule precursors, and $m_c$ is the mass of a
single chondrule precursor.
The stopping time for chondrule precursors can be estimated from the Epstein
gas drag law \citep{Epstein1924}
\begin{equation}\label{eq:ts}
    t_s = \frac{\rho_d a_d}{\rho_g c_s},
\end{equation}
where $\rho_d$ and $a_d$ are the material density and radius of the chondrule precursor.
$t_s$ increases over time as the gas depletes and $\rho_g$ drops in our disk
model. For the time range of chondrule formation (Figure \ref{fig:probability}),
$t_s$ of mm-sized particles in the disk mid-plane  
within $3.5~\mathrm{AU}$ is less than 1\% of the orbital time.
Therefore it is safe to assume that the
chondrule precursors are well-coupled with the gas, and $v_\mathrm{rel}$ is
simply the relative velocity between the planetesimal and the gas.
\citet{Morris2012} simulated the bow shocks from $r_p > 500~\mathrm{km}$
planetesimals, and found that the cross-section of the shock region is roughly
proportional to the geometric cross-section $\sigma_c \approx \pi (1.6 r_p)^2$.
For smaller planetesimals, the
gravitational focusing from the planetesimals is weaker and their atmosphere is
thinner, and $\sigma_c$ approaches the geometric cross-section $\pi r_p^2$.
Since the exact dependence of $\sigma_c$ on $r_p$ is unknown,
we adopt $\sigma_c = \pi (1.6 r_p)^2$ for planetesimals of
all sizes in our simulations for simplicity.
We assume that the chondrule precursors are distributed in a dust disk with a 
similar profile as the gas disk (Equation (\ref{eq:rho_g_z})), 
$\rho_c(R, z)=\rho_c(R,0)\exp(-z^2/2H_c^2)$. $\Sigma_c$ is the surface density
of chondrule precursors and follows the same scaling with radius as the MMSN, 
$\Sigma_c \propto R^{-1.5}$. $H_c$ is the scale height of the chondrule
precursors from the disk mid-plane. We assume 
the planetesimals to be confined near the mid-plane of the disk, and use the
mid-plane density of the chondrule precursors 
$\rho_c = \Sigma_c/(\sqrt{2\pi} H_c)$ in Equation (\ref{eq:M_c}). 
Due to gas drag and the gravity from the central star,
dust grains settle down toward the disk mid-plane 
and thus have a smaller scale height than the gas disk. The dust disk
scale-height depends both on the grain size which determines the gas drag, and
the turbulence level in the disk which affects how efficiently grains can be lifted
away from the mid-plane. We adopt the dust disk
scale-height $H_c$ from numerical simulations of turbulent disk
with magnetic fields and ambipolar diffusion by \citet{Xu2017}:
\begin{equation}
    H_c = \sqrt{\frac{\alpha_z}{\tau_s}} H,
\end{equation}
where $\alpha_z=7.8\times 10^{-4}$ is the vertical turbulence diffusion
coefficient, and $H$ the gas disk scale-height.\footnote{The simulations in 
\citet{Xu2017} adopt an ambipolar diffusion 
Elsasser number $Am=1$, which is more suitable for the
outer disk $\gtrsim 30~\mathrm{AU}$. In the inner disk where chondrules form,
the ambipolar diffusion and other non-ideal MHD effects are likely to be stronger,
and therefore the turbulence in the disk is likely to be even weaker. This will
lead to a smaller $H_c$ and a higher chondrule formation efficiency.}
$\tau_s=\Omega_k t_s$ is the dimensionless stopping time.
As the disk depletes over time, 
the gas density $\rho_g$ drops, leading to larger $\tau_s$ and smaller $H_c$.
We choose $\rho_d=1~\mathrm{g/cm^3}$
and $a_d=1~\mathrm{mm}$ for the chondrule precursors, which gives $H_c/H=0.2-0.9$
between $1.5-3.0~\mathrm{AU}$ and $t/\tau_\mathrm{dep}=2.5-4.5$. 

Let $P_c$ be the average probability that a particular dust grain between disk
radius $R_1$ and $R_2$ becomes a chondrule. Then the contribution to $P_c$ from
a planetesimal between $R_1$ and $R_2$ is given by
\begin{align}\label{eq:P_c}
    \frac{\di P_c}{\di t} &= 
    \frac{1}{M_{c, \mathrm{tot}}}\frac{\di M_{c, \mathrm{heating}}}{\di t}
    = \frac{\sigma_c v_\mathrm{rel}\Sigma_c(R)/(\sqrt{2\pi}H_c(R))}{\int_{R_1}^{R_2}
    \Sigma_c(R') 2\pi R' \di R'}\\
    &=\frac{\sigma_c v_\mathrm{rel} R^{-3/2}}{4\pi \sqrt{2\pi} H_c (R_2^{1/2} -
    R_1^{1/2})}, \nonumber
\end{align}
where $M_{c,\mathrm{tot}}$ is the total mass of chondrule precursors.
We choose the chondrule forming region to be between $R_2 = 3.0~\mathrm{AU}$
and $R_1=1.5~\mathrm{AU}$. 

\begin{figure}[!ht]
\centering
\includegraphics[width=\linewidth]{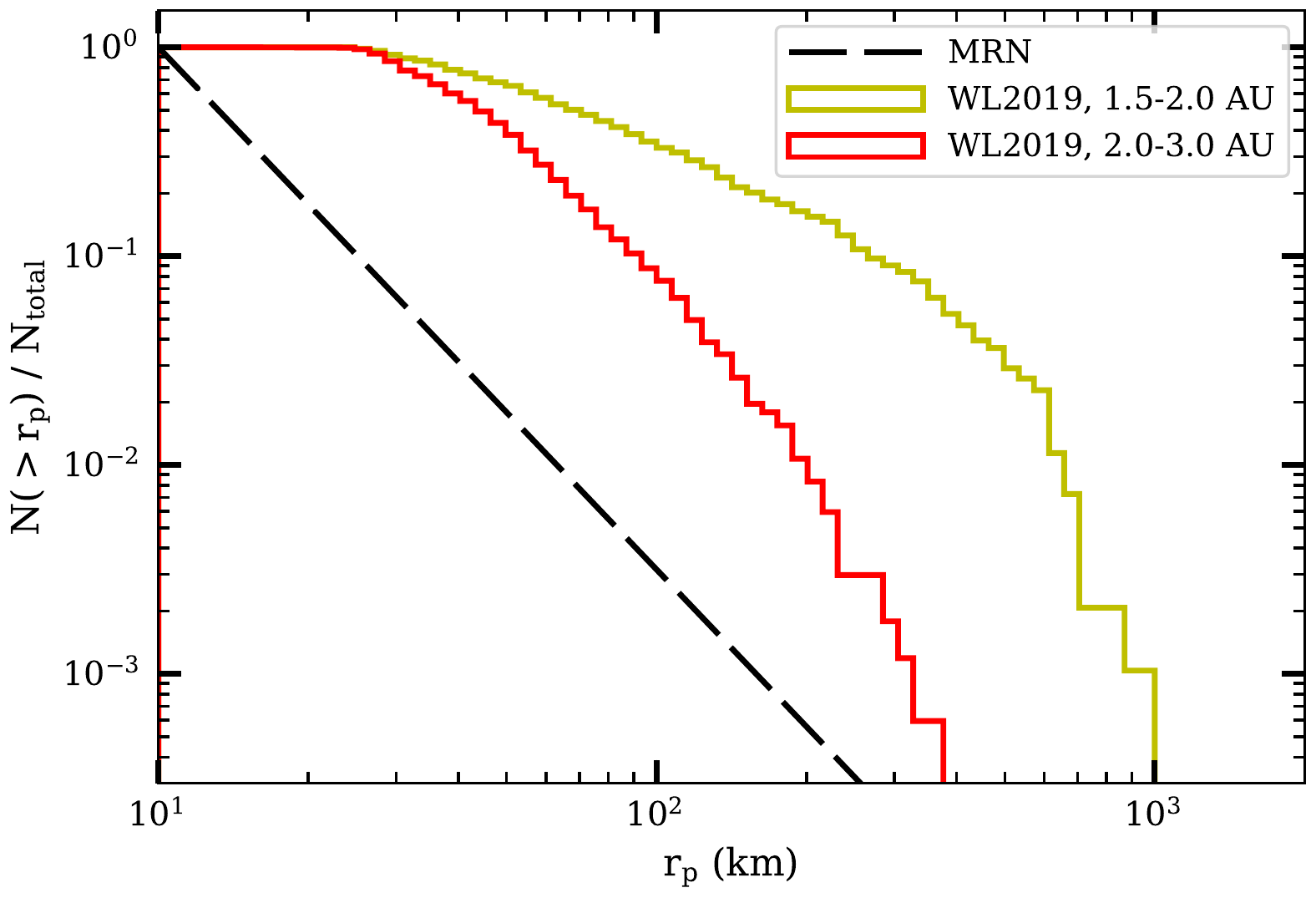}
    \caption{The cumulative
    size distribution of planetesimals from the collisional growth
    simulations by \citetalias{WL2019} at $1~\mathrm{Myr}$
    in different disk radius ranges (see legends),
    and the standard MRN size distribution of $\di N/\di r_p \propto r_p^{-3.5}$ 
    \citep{MRN1977}. In the MRN distrubtion, we set the largest planetesimal
    size to be $r_p=2000~\mathrm{km}$.
    In \citetalias{WL2019}, the largest planetesimal beyond
    $1.5~\mathrm{AU}$ has grown to $r_p \approx 1000~\mathrm{km}$ at $1~\mathrm{Myr}$.
    \label{fig:rp_dist}
}
\end{figure}

The contribution to $P_c$ from one planetesimal is calculated by integrating
Equation (\ref{eq:P_c}) over the time period when
$v_\mathrm{rel}$ is in the suitable range for chondrule heating.
With many planetesimals in our simulations, we can then estimate the average
contribution to
$P_c$ from planetesimals of different sizes that satisfy the chondrule
formation criteria, as well as the distribution of
$P_c$ over physical parameters such as time, relative velocity, gas density,
and radial locations in the disk. 
We adopt the size
distribution of planetesimals from collisional evolution simulations by
\citet[][hereafter \citetalias{WL2019}]{WL2019}.
In their simulation, planetesimals in the inner disk grow
faster than in
the outer disk due to the higher surface density of solids. Because
the first chondrules formed about $1~\mathrm{Myr}$ after the CAIs \citep{Scott2007},
we use the simulated size distribution of planetesimals from
\citetalias{WL2019} at $1~\mathrm{Myr}$, as shown in Figure \ref{fig:rp_dist}.
In their simulations, the planetesimals located beyond $1.5~\mathrm{AU}$ have not
had enough time to reach the runaway or oligarchic growth stage at
$1~\mathrm{Myr}$ due to collisional fragmentation.
To investigate the dependence of chondrule
formation on the size distribution of planetesimals, we also 
adopt a standard power-law size distribution of the planetesimals for $r_p =
10- 2000~\mathrm{km}$, 
$\di N/\di r_p \propto r_p^{-3.5}$ \citep[][hereafter MRN]{MRN1977} for
comparison. As shown in Section
\ref{section:emax}, planetesimals with $r_p \gtrsim 2000~\mathrm{km}$
do not make a significant contribution to chondrule heating due to the strong
tidal force from the gas disk. The surface density of the planetesimals
is assumed to be 1\% of the initial surface density of the gas disk.  

\section{Results}\label{section:results}
\subsection{Orbital Evolution of Planetesimals\label{sec:result:orbit}}
\begin{figure*}[!ht]
\centering
\includegraphics[width=\linewidth]{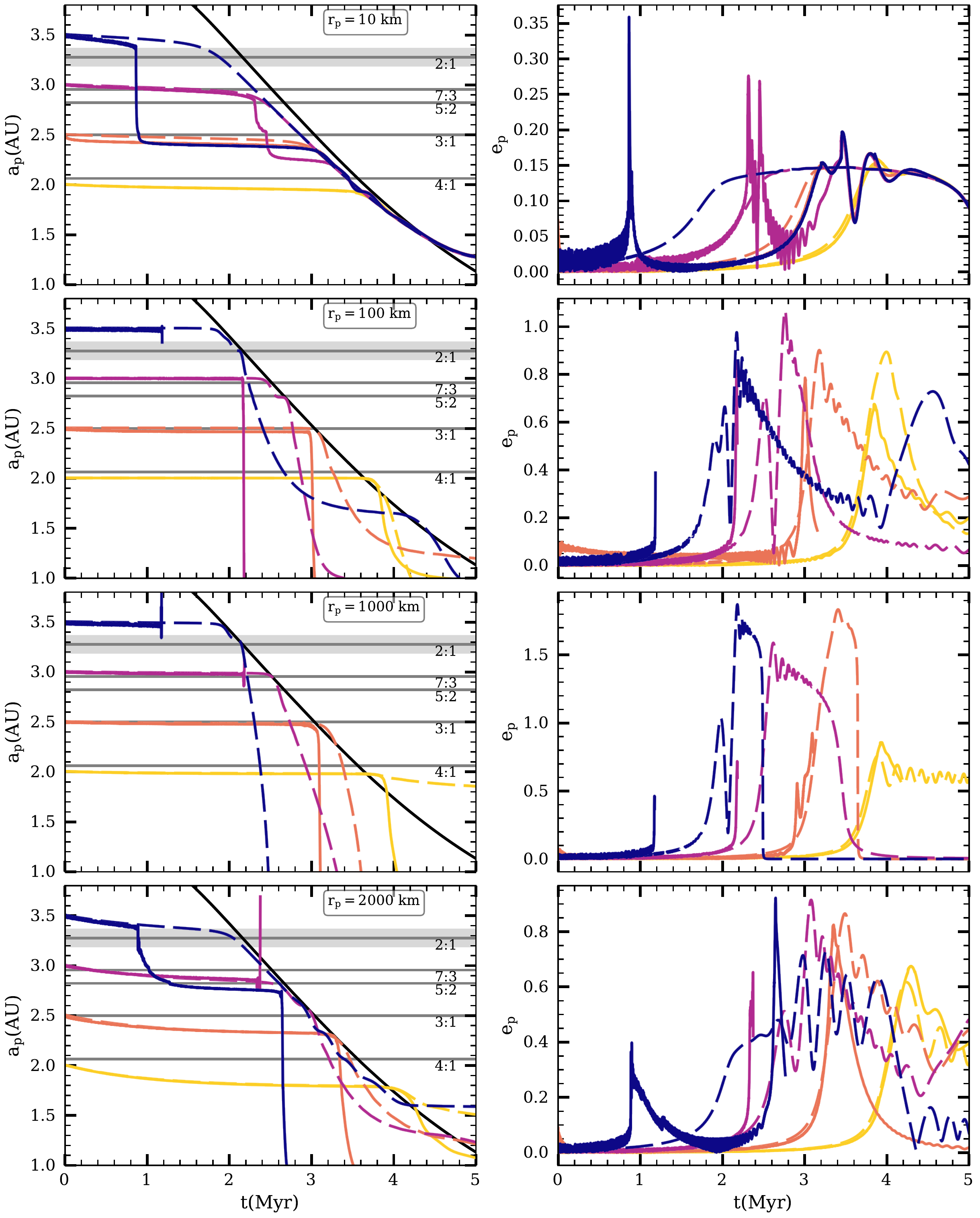}
    \caption{Semi-major axis (left) and eccentricity (right) evolution of
    planetesimals with radius from $10$ to $2000~\mathrm{km}$ (top to
    bottom). The black solid lines in the left panels show the location
    of the $\nu_5$ resonance, same as the black solid line in Figure \ref{fig:nu5}.
    Dashed and
    solid lines show the results from semi-analytic estimation and numerical
    simulations. Different colors represent planetesimals with different
    initial semi-major axes $a_{p,0}=2.0, ~2.5, ~3.0$ and $3.5~\mathrm{AU}$
    (see left panels at $t=0$).
    The horizontal grey lines mark the locations of mean motion
    resonances. The gray shaded region indicates the width of the 2:1 mean motion
    resonance for a planetesimal with $e_p = 0.2$.
    \label{fig:orbits}
}
\end{figure*}

The time-evolution of the semimajor-axis and eccentricity of individual
planetesimals in the fiducial model
is shown in Figure \ref{fig:orbits}. The sizes of planetesimals
are selected to represent different regimes of gas-damping (see Section
\ref{section:gas-damping}):
$r_p=10~\mathrm{km}$ represents the ``strong aerodynamic damping'' group,
$r_p=100~\mathrm{km}$ and $r_p=1000~\mathrm{km}$ the ``weakly coupled''
group, and
$r_p=2000~\mathrm{km}$ the ``strong tidal damping'' group. Planetesimals in
each group show distinctively different behaviors.

Before we delve into the details of the orbital evolution of
planetesimals, it is helpful to understand the comparison between the
semi-analytic results and numerical simulations. There are two major
differences. First, the mean-motion resonances are not included in the
semi-analytic calculations. The 2:1 mean-motion resonance with Jupiter at
$3.3~\mathrm{AU}$ is
especially powerful in eccentricity excitation of planetesimals. The 2:1
resonance also has a large width, which depends on the eccentricity of the
planetesimal, and is very wide at low eccentricity (see \citet{MD1999} Chapter
8.7). The planetesimals with initial semi-major axes at
$3.5~\mathrm{AU}$ start within the width of 2:1 resonance. Their eccentricities
increase, leading to the inward migration by gas drag which brings them closer to
the 2:1 resonance, resulting in even stronger eccentricity excitation. As a
result, many of them gain very large eccentricities, and some are ejected from
the system by close encounters with Jupiter. Similarly, planetesimals with
initial semi-major axes at $3.0~\mathrm{AU}$ are also influenced by the 7:3,
5:2 and 3:1 resonances, although these resonances are weaker than the 2:1
resonance. Second, because our semi-analytic treatment of the secular resonance
is only valid to the first order in eccentricity,
it is no longer a good approximation when $e_p$ is close to unity. For example,
planetesimals with $a_{p, 0}=2.0~\mathrm{AU}$ in Figure \ref{fig:orbits}
(yellow lines) are not strongly affected by the mean motion resonances,
and the semi-analytic models and numerical simulations agree 
well when $e_p \lesssim 0.5$, confirming that the numerical simulations 
correctly capture the secular resonance. At $e_p \gtrsim 0.5$, the
semi-analytic solutions deviate from the numerical simulations due to the
failure of the first-order approximation. 

The orbital evolution of the $10~\mathrm{km}$-sized planetesimals is shown in the
top panels of Figure \ref{fig:orbits}. Planetesimals with $a_{p,
0}=3.0~\mathrm{AU}$ and $3.5~\mathrm{AU}$ are excited to moderate
eccentricities $e_p\sim 0.3$ by mean-motion resonance and quickly migrate
inward due to the
strong aerodynamic gas drag. When caught in the secular resonance, they can
``surf'' with the resonance: they maintain small eccentricities $e_p \sim 0.2$,
which allow them to migrate inward along with the $\nu_5$. 

Planetesimals with sizes $r_p = 100$ and $1000~\mathrm{km}$ experience much
weaker gas drag (middle panels of Figure \ref{fig:orbits}). 
As a result, their eccentricities can be easily excited to
high values close to unity. Note that the eccentricity is not limited to be
below unity in the semi-analytic calculations, but the first order
approximation is also no longer valid in this case.
Most of the planetesimals quickly leave the simulation domain 
either by rapid inward migration or close-encounter with
Jupiter. As will be discussed in Section \ref{section:probability}, chondrules can 
be heated efficiently in the bow shock of
this ``weakly coupled'' group of planetesimals.

Very large planetesimals with $r_p \gtrsim 2000~\mathrm{km}$ are affected by
strong tidal damping (lower panels of Figure \ref{fig:orbits}). Unlike 
the aerodynamic drag, tidal drag can cause inward migration
even when the planetesimal is in a circular orbit.
This can be seen from Equations (\ref{eq:dapdt_aero}) and 
(\ref{eq:dapdt_tidal}): for a circular orbit ($e_p=0$), 
the timescale for inward-migration due to aerodynamic drag is proportional to
$\eta^2$ but the timescale for tidal drag is proportional to $\eta$. The 
inward migration of $2000~\mathrm{km}$ planetesimals causes them to
encounter the secular resonance later at smaller disk radii. Planetesimals with
$r_p \gtrsim 4000~\mathrm{km}$ migrate inward so rapidly that they never
encounter the sweeping secular resonance (see Figure \ref{fig:emax} and
discussion).

\subsection{Size-dependent Eccentricity Excitation\label{section:emax}}
\begin{figure}[!ht]
\centering
\includegraphics[width=\linewidth]{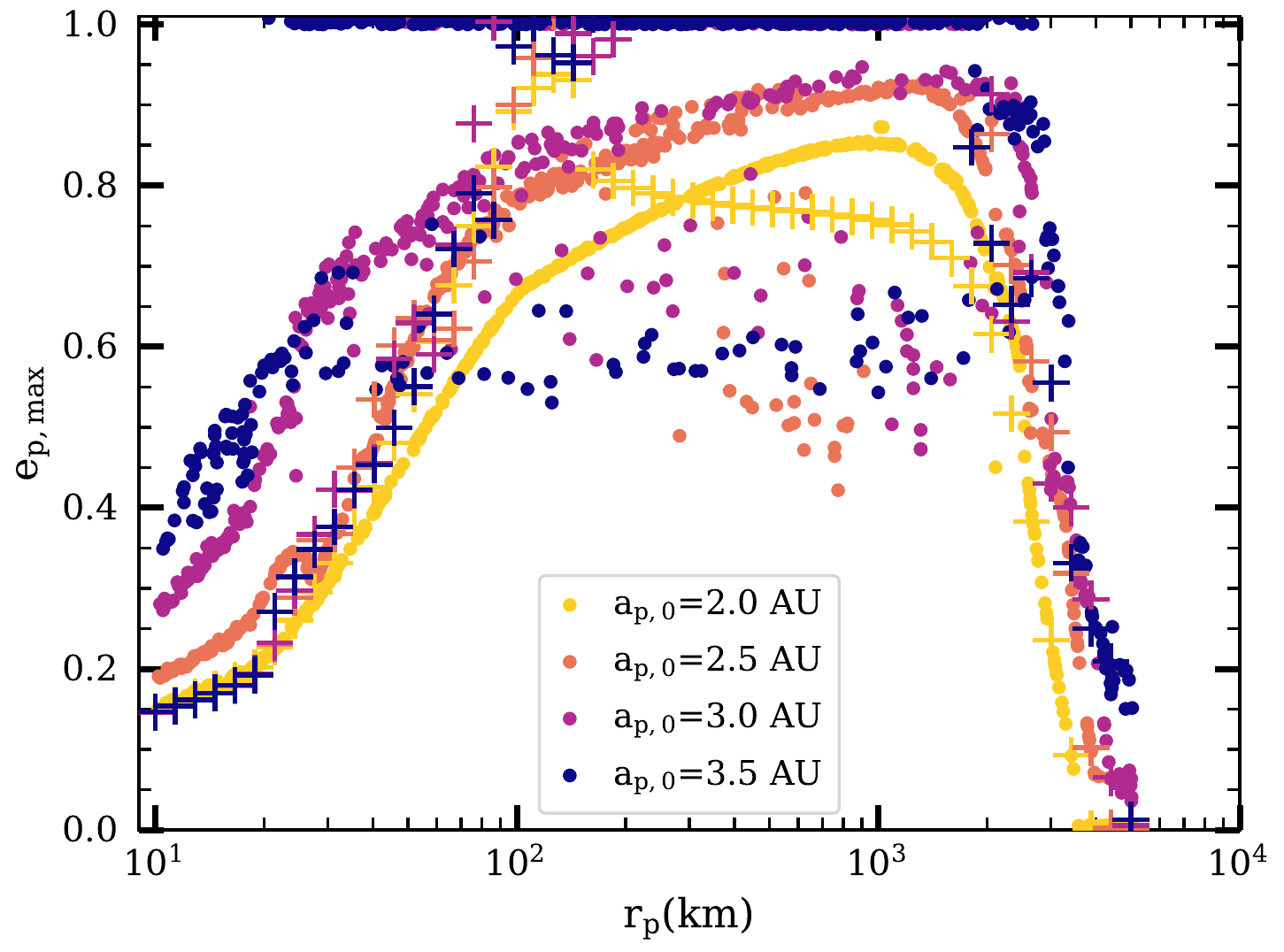}
\caption{Maximum eccentricities versus radii of planetesimals for
semi-analytic models (crosses) and numerical simulations (circles). Each dot
represents one planetesimal. Different colors
represent planetesimals with different initial semi-major axes  
    $a_{p,0}=2.0,~2.5,~3.0$ and $3.5~\mathrm{AU}$ (same as Figure
    \ref{fig:orbits}, see also the legend). 
The dots with $e_p=1$ represent planetesimals that are ejected from the system.}
\label{fig:emax}
\end{figure}
To summarise the size-dependent eccentricity excitation of planetesimals, we
plot the maximum eccentricity of each planetesimal versus its size in the
fiducial model in Figure \ref{fig:emax}. 
The semi-analytic models and numerical simulations give
similar results, except for very high eccentricities $e_p \gtrsim 0.5$ when the
semi-analytic approximation is no longer valid, and for
planetesimals with initial semi-major axes $a_{p,0}\geq 3~\mathrm{AU}$ (blue
and purple symbols) that are strongly affected by mean-motion resonances. 
At a fixed $r_p$ and $a_{p,0}$, the spread of the $e_{p, \mathrm{max}}$ in
numerical simulations is caused by the random initial phase angles. In the
``strongly coupled'' group of small and large planetesimals, the spread of
$e_{p, \mathrm{max}}$ is small at a given $a_{p,0}$, showing that their orbital
evolution is not sensitive to the initial phase angle. In the ``weakly
coupled'' group of intermediate size planetesimals, most of the planetesimals
are still in the main distribution of $e_{p, \mathrm{max}}$ as a function of
$r_p$. However, there are some outliers that have lower eccentricity $e_p\sim
0.6$ or extremely high eccentricity $e_p \sim 1$. These outliers are
planetesimals that have close encounters with Jupiter, and planetesimals that
pass through the $\nu_5$ resonance with less eccentricity
excitation or inward migration. \citet{Zheng2017} discusses this in more
details (see their Figure 9).

The dependence of the planetesimal's maximum eccentricity on its size exhibits
the behavior one would expect from the gas drag laws discussed 
in Section \ref{section:gas-damping}. As the size
increases, the maximum eccentricity $e_{p, \mathrm{max}}$ 
first increases due to decreased aerodynamic
drag and then decreases again due to increased tidal drag. Because the
aerodynamic damping timescale is proportional to $r_p$ and the tidal damping
timescale is proportional to $r_p^{-3}$ (Equations (\ref{eq:tau_aero}) and
(\ref{eq:tau_tidal})), the drop of $e_{p, \mathrm{max}}$ is steeper at large
$r_p$.  Very large planetesimals with $r_p \gtrsim 4000~\mathrm{km}$
migrate inward rapidly due to the tidal drag, and never
encounter the secular resonance. Planetesimals with $r_p \approx
50-2000~\mathrm{km}$ can be excited to very high eccentricities $e_p > 0.7$
and are the primary contributors to chondrule formation. 

\subsection{Probability of Chondrule Formation\label{section:probability}}
\subsubsection{Overall Behavior\label{section:overall}}
\begin{table}[htbp]
    \caption{Chondrule Formation Probability and Age Spread\tablenotemark{a}}
    \label{table:Pc}
    \centering
    \begin{tabular}{c cc cc}
        \tableline
        \tableline
         &\multicolumn{2}{c}{\citetalias{WL2019}} 
         &\multicolumn{2}{c}{MRN} \\
        model ID &$P_{c, \mathrm{tot}}$ &$\Delta t / \tau_\mathrm{dep}$
        &$P_{c, \mathrm{tot}}$ &$\Delta t / \tau_\mathrm{dep}$ \\
        \tableline
        fiducial &8.1\% &1.3 &4.2\% &1.1\\
        DEP0p5   &3.9\% &1.4 &1.5\% &1.4\\
        DEP2     &7.2\% &1.7 &3.6\% &1.2\\
        TDL5     &8.9\% &1.6 &4.2\% &1.2\\
        EJ0p1    &6.6\% &1.3 &3.9\% &1.2\\
        ST       &8.0\% &1.8 &4.2\% &1.5\\
        STR      &5.7\% &1.8 &3.1\% &1.8\\
        \tableline
        \tableline
    \tablenotetext{1}{Total chondrule formation probability $P_{c,
        \mathrm{tot}}$ and the age spread of chondrules $\Delta t$ normalized
        by the disk depletion time $\tau_\mathrm{dep}$ for \citetalias{WL2019}
        and MRN planetesimal size distribution. $\Delta t$ is defined as the
        time interval within which 70\% of chondrules are formed (see also
        Figure \ref{fig:probability}).}
    \end{tabular}
\end{table}

The total chondrule formation probability and the age spread of chondrules
in different numerical models are summarized in Table \ref{table:Pc}. The age
spread of chondrules is $\sim 1-2$ times the disk depletion time $\tau_\mathrm{dep}$
in all of our models. This is because most chondrules form when the secular
resonance sweeps through the main asteroid belt region,
which happens on a timescale comparable to the disk depletion time. The
total chondrule formation probability $P_{c, \mathrm{tot}}$
is about twice higher in the case of
\citetalias{WL2019} size distribution compared to the MRN size distribution.
This is because there is a larger number of smaller planetesimals in the
``weakly coupled'' group of $r_p \sim 50-2000~\mathrm{km}$ for the
\citetalias{WL2019} size distribution (see Figure \ref{fig:rp_dist}). For a
given total mass of planetesimals, smaller planetesimals have a larger
surface area, and $P_c$ is proportional to the
surface area that the planetesimals' bow shocks sweep through. Indeed, we
find that the total surface area of planetesimals from the \citetalias{WL2019}
size distribution is about twice that from the MRN size distribution.

\begin{figure*}[!ht]
\centering
\includegraphics[width=\linewidth]{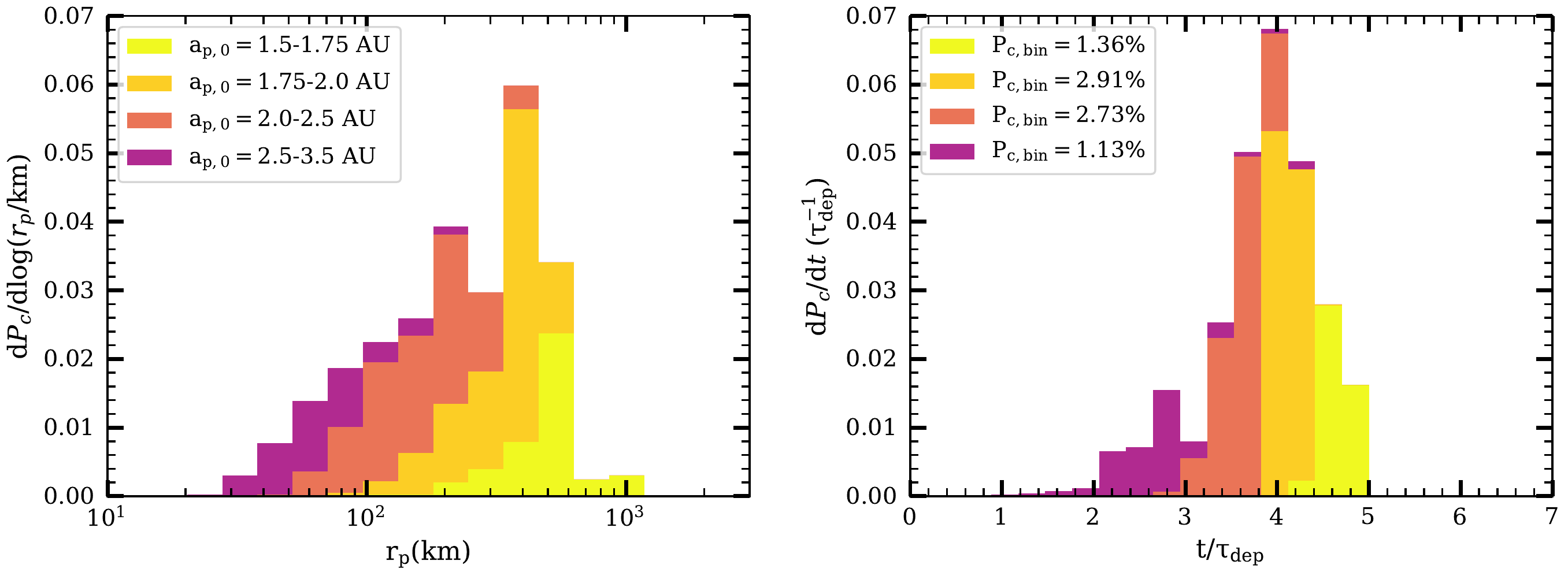}
    \caption{Stacked histograms of the chondrule formation probability in the
    fiducial model.
    {\sl Left panel:} The contribution to the total
    chondrule formation probability $P_c$ 
    made by planetesimals of different sizes at each
    initial semi-major axes bin (see figure legends).
    The planetesimals  
    are assumed to have a surface density of 1\% of the initial gas disk mass, 
    and a size distribution from
    simulations by \citetalias{WL2019} (see Figure \ref{fig:rp_dist} and
    discussions).
    {\sl Right panel}: Similar to the left panel, but 
    for the contribution to $P_c$ at different times. 
    The total chondrule formation probability in each $a_{p,0}$ bin,
    $P_{c,bin}$, is marked in the figure legends.}
\label{fig:probability}
\end{figure*}

\begin{figure*}[htbp]
\centering
\includegraphics[width=\linewidth]{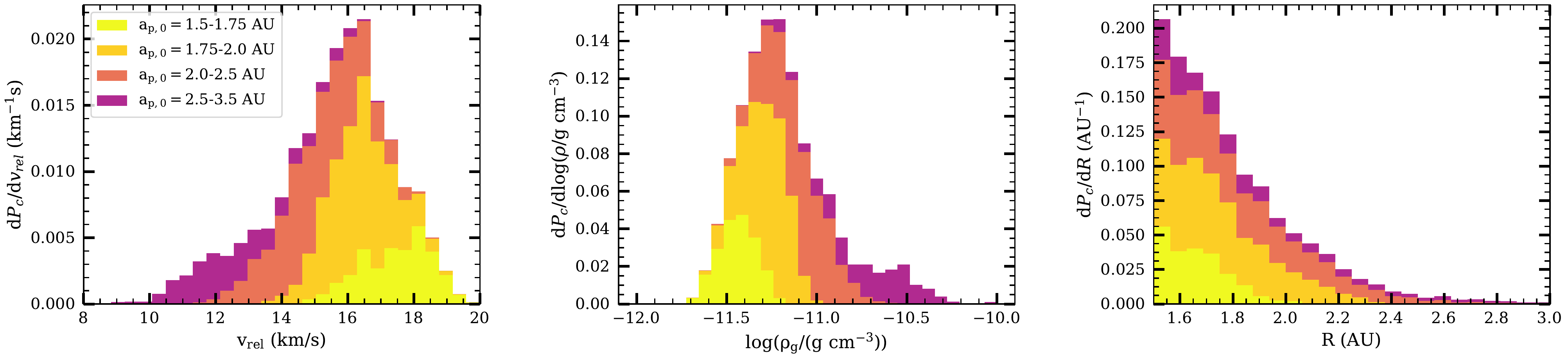}
\caption{Similar to the right panel of Figure \ref{fig:probability}, but for
    the contribution to 
$P_c$ by planetesimals with different relative velocities with respect to
the gas ({\sl left}), and by chondrule heating events with different local gas densities
    ({\sl middle}) and distances from the Sun ({\sl right}).}
\label{fig:probability_details}
\end{figure*}

The contribution to the total chondrule formation probability $P_c$ made by
planetesimals of different sizes and at different times in the fiducial
model is plotted in Figure \ref{fig:probability}. The total stacked
histograms show the distribution of $P_c$ as a function of planetesimal size.
The contributions to $P_c$ by planetesimals from different
initial semi-major axis bins are indicated by different colors. It is evident that 
planetesimals with sizes $\sim 50-1000~\mathrm{km}$ are the main
contributors to chondrule formation.
The \citetalias{WL2019} size distribution does not have
planetesimals with $r_p\gtrsim 1000~\mathrm{km}$, but in the case of the MRN
size distribution, $1000-2000~\mathrm{km}$ planetesimals also contribute
to chondrule formation. This is expected, as the $\sim 50-2000~\mathrm{km}$
planetesimals are in the ``weakly coupled''
group that can be excited to large eccentricities. 

Overall, chondrule formation are dominated by planetesimals with initial
semi-major axis $1.5~\mathrm{AU} \lesssim a_{p,0} \lesssim 2.5~\mathrm{AU}$.
Because the Keplerian velocity is smaller further away from the Sun,
it is difficult for planetesimals with $a_{p,0} \gtrsim 2.5~\mathrm{AU}$ to
gain high enough velocities for chondrule formation.
This can be seen in the distribution of $P_c$ with relative velocity
in the left panel of Figure \ref{fig:probability_details}. Planetesimals with 
$a_{p,0} \lesssim 1.5~\mathrm{AU}$ do not contribute significantly to
chondrule formation, because it is less likely for their orbits to cross the
asteroid region of $1.5-3.0~\mathrm{AU}$, where we assume chondrules are
formed. In principle, chondrules
can form inside $1.5~\mathrm{AU}$ and be transported outward to the asteroid
belt region today, but this scenario is beyond the scope of this paper.

Generally speaking from Figure \ref{fig:probability_details},
most chondrules form in shocks of very high
velocities $v_\mathrm{rel}\approx 12-18~\mathrm{km/s}$, in low density gas
$\rho_g\approx 10^{-11.5}-10^{-10.5}~\mathrm{g/cm^3}$ when the disk is
significantly depleted relative to the MMSN, and in the inner
asteroid belt region $R\lesssim 2.5~\mathrm{AU}$. 

Mean motion resonance, on the other hand, does not cause a significant amount of
chondrule heating. Although the planetesimals close to the 2:1 resonance with
Jupiter can be excited to eccentricities $e_p>0.5$, they contribute
very little to chondrule heating (Figure \ref{fig:probability}). This is because
the 2:1 resonance is in the outer region of the disk where the Keplerian
velocity is relatively low, and $v_\mathrm{rel}$ is hardly ever large enough for
chondrule formation. This is consistent with the result in \citet{Nagasawa2014}. 

\subsubsection{Dependence on Model Parameters}
Table \ref{table:Pc} shows that the total chondrule formation probability
$P_{c,\mathrm{tot}}\approx 4-9\%$, only varies by a factor of $\sim 2$ for a wide
range of model parameters. In this section, we discuss the reasons why
$P_{c,\mathrm{tot}}$ is relatively insensitive to these parameters.

Doubling the disk depletion timescale to $\tau_\mathrm{dep}=2~\mathrm{Myr}$
(model DEP2) does not have a significant effect on the chondrule
formation probability. This is because both the location of $\nu_5$ and the
disk mass only depend on $t/\tau_\mathrm{dep}$. A planetesimal with initial
semi-major axis $a_{p,0}$ encounters $\nu_5$ at a certain value of
$t/\tau_\mathrm{dep}$ and a corresponding disk mass, 
independent of $\tau_\mathrm{dep}$ (Figure \ref{fig:nu5}). 
Therefore, the eccentricity excitation and damping rates of the
planetesimal are largely unchanged. If $\tau_\mathrm{dep}$ is too short, however,
$\nu_5$ migrates in so rapidly that the planetesimals do not have enough time
to react to the $\nu_5$ resonance before it passes through, leading to less
efficient eccentricity excitation. This is why $P_{c,\mathrm{tot}}$ in model
DEP0p5 is reduced.  

Moreover, $P_{c,\mathrm{tot}}$ is rather 
insensitive to the initial eccentricity of
Jupiter (model EJ0p1) and the tidal damping timescale (model TDL5). The
planetesimal's eccentricity excitation rate is proportional to $e_J$
(Equation \ref{eq:depJdt}), and its eccentricity tidal damping rate is
inversely proportional to $\tau_\mathrm{tidal}$ (Equation
(\ref{eq:depdt_tidal})). However, the chondrule formation probability is not
sensitive to the eccentricity excitation and damping rates, as long as the
planetesimal can gain high enough eccentricity to cause chondrule formation.
When the eccentricity of the planetesimal is higher,
although the chondrule formation rate is higher,
the planetesimal also migrates inward faster and thus spends less time in the
asteroid belt region. These two effects roughly cancel each other, and the resulting
$P_{c,\mathrm{tot}}$ is largely unchanged.

Introducing Saturn does not have a strong effect on $P_{c,\mathrm{tot}}$ either
(models ST and STR). The sweeping secular resonance by Saturn
occurs at a later time than that by Jupiter. As most planetesimals in the
``weakly damping'' group move inward rapidly
when they encounter the Jovian secular resonance and gain high eccentricities,
they do not encounter the secular resonance by Saturn most of the time. 
Saturn modulates the eccentricity of Jupiter, changing the eccentricity
excitation rate of the planetesimals. Saturn's
gravity and the larger gap size also change the precession rate of Jupiter, and
cause the $\nu_5$ resonance to occur at a slightly different time when the disk
mass is different, changing the eccentricity damping rate of planetesimals.
However, $P_{c,\mathrm{tot}}$ is relatively 
insensitive to the eccentricity excitation rate and damping rate for the
reasons stated in the previous paragraph. 

\subsubsection{Comparison to Observations}
Can the chondrule formation probability in our model explain the fraction of
chondrules in our current asteroid belt? 
From meteorite samples found on Earth,
ordinary chondrites are the most commonly found type of meteorites \citep{SD1988}. 
Chondrules make up $60-80$\% of the volume of ordinary chondrites, and a rough
estimate gives that $\sim 10$\% of the present day asteroid belt can be made of
chondrules \citep{Desch2005}. However, there are also arguments that
chondrules may be much rarer than commonly believed. 
\citet{Sears1998} pointed out two main sources of bias: 
First, ordinary chondrites are much more robust than volatile-rich
carbonaceous chondrites, and
thus much more likely to survive falling through the Earth's atmosphere. Second,
the meteorites that fall on Earth may not be a fair sample of the
asteroid belt. He estimated that after the correction of biases, 
ordinary chondrites
make up less than $4$\% of the asteroid belt. \citet{MC1999} argues
that the sample of extraterrestrial dust particles is much less biased, and
less than $\sim 1$\% of them have a composition similar to ordinary chondrites. 
In general, our simulated chondrule formation probability of $4-9\%$ is
consistent with these observational constraints.

\section{Discussion}\label{section:discussions}
\subsection{The Depletion Timescale of the Solar Nebula}
Assuming the ages of chondrules measured from radioactive isotopes indicate the
time that they went through flash-heating, 
the age spread of chondrules can be indicative of the disk depletion timescale,
as shown in Table \ref{table:Pc} and discussed in Section
\ref{section:overall}.

Isotopic measurement of chondrules' ages indicate that chondrule formation
started 0.7 Myr after the formation of CAIs, and lasted for at least 2.4 Myr
and potentially up to 5 Myr, with most chondrules formed
$1-3~\mathrm{Myr}$ after the CAIs
\citep[see reviews by][]{Connolly2006, Scott2007}.
This is consistent with a disk depletion time of
$\tau_\mathrm{dep}\approx 1~\mathrm{Myr}$. This is also broadly consistent with the 
observations of young stellar clusters, which suggest that about 
half of the young stars lose their disk in $\sim 1-3~\mathrm{Myr}$
\citep{Haisch2001,Armitage2003,Ribas2015,Richert2018}.

We note that $t=0$ in Figures \ref{fig:probability}
does not indicate the time of CAI formation. It
corresponds to the time when the disk mass equals the MMSN mass, which is
chosen to be the initial condition of our models (see also Figure
\ref{fig:story}). CAI formation can occur
either before or after $t=0$ in our models. Moreover, the relevant
timescale for chondrule formation is measured by
the migration timescale of $\nu_5$, or $\tau_\mathrm{dep}$, which does not
necessarily correspond to a physical time.
There is no reason why $\tau_\mathrm{dep}$ in Equation
(\ref{eq:f_g}) has to remain constant throughout the disk evolution. 
Gap formation may actually lead to an accumulation of disk gas in the region
outside Jupiter’s orbit while the inner disk may deplete rapidly. 
Photoevaporation may also accelerate the pace of surface density decline 
in the inner disk region. However, the conclusion that the age spread of
chondrules is comparable to the time that $\nu_5$ sweeps through the asteroid
belt region is unchanged even if $\tau_\mathrm{dep}$ varies with time.

\subsection{The Formation of Jupiter}
In our model, the formation of Jupiter is essential for chondrule heating.
However, unlike the model of chondrule heating by mean motion resonance
\citep{Weidenschilling1998}, the onset of chondrule heating and Jupiter's
formation do not have to be coeval. The only constraint on the formation time
of Jupiter is that it has to be before the formation of chondrules, not more
than $0.7~\mathrm{Myr}$ after the CAIs formed.
The timing of chondrule formation indicates when
the $\nu_5$ resonance passes through the asteroid belt, which is directly tied
to a disk mass of $1-10\%$ of the MMSN. A recent study by \citet{Kruijer2017} 
suggests that the composition of
different types of meteorites indicates that Jupiter formed 
and opened a gap in the disk within $1~\mathrm{Myr}$ after the formation of the CAIs.

Given that Jupiter is likely to form in the gas-rich disk before the
formation of chondrules, the sweeping secular resonance seems to be hardly
avoidable: as long as the gas disk is more than $\sim 10\%$ of the MMSN after
the formation of Jupiter, the
$\nu_5$ resonance will pass through the asteroid belt region and excite the
eccentricities of any large planetesimals that may be there. There is indeed 
evidence that large planetesimals were present before chondrule formation: 
iron meteorites, which come from the fragments of cores of the earliest-formed
differentiated planetesimals, 
are dated to be accreted as early as $0.1-0.3~\mathrm{Myr}$ after the
formation of CAIs \citep{Kruijer2014, Kruijer2017}. 

In addition, the formation of Jupiter can lead to relocation of
nearby planetesimals through scattering \citep{ZL2007}. 
The subsequent propagation of Jupiter’s sweeping secular resonance can 
lead to collisional fragmentation, merger, and effective clearing of 
the main belt region \citep{NLT2005, TNL2008, Zheng:submitted}.

\section{Summary}\label{section:summary}
In this paper, we propose that chondrules can form efficiently when the
Jovian sweeping secular resonance passes through the asteroid belt region and
excites the eccentricities of planetesimals in the early solar protoplanetary
disk. We use semi-analytic models and numerical simulations to study the
orbital evolution of planetesimals and its effect on chondrule formation.
Our main findings are summarized as follows.
\begin{enumerate}
    \item Planetesimals with sizes $\sim 50-2000~\mathrm{km}$ are subject to
        relatively weak gas drag in the disk. They can be excited to eccentricities 
        $e_p > 0.6$ by the Jovian sweeping secular resonance in the asteroid
        belt region and cause chondrule formation. Smaller or bigger
        planetesimals suffer from either strong eccentricity damping or rapid inward
        migration by the gas drag, 
        and thus cannot gain high enough eccentricities required for
        chondrule heating (Figure \ref{fig:emax}).
    \item Most chondrules form in high
        velocity shocks $v_\mathrm{rel}\approx 12-18~\mathrm{km/s}$, 
        in low density gas $\rho_g\approx 10^{-11.5}-10^{-10.5}~\mathrm{g/cm^3}$ 
        when the disk is depleted to $1-10\%$ of the mass of
        the MMSN, and in the inner asteroid belt region $R\lesssim
        2.5~\mathrm{AU}$ (Figure \ref{fig:probability_details}).
    \item The average chondrule formation probability between
        $1.5-3.0~\mathrm{AU}$ is about $4-9\%$, 
        consistent with observational constraints
        (Table \ref{table:Pc}).
    \item Our model suggests that the depletion timescale for the protoplanetary
        disk around the Sun is comparable to the age spread of chondrules at
        $\tau_\mathrm{dep}\approx 1~\mathrm{Myr}$ (Table \ref{table:Pc}), and 
        that Jupiter must have formed before the formation
        of chondrules, not more than $0.7~\mathrm{Myr}$ after the CAIs.
\end{enumerate}

\section{Acknowledgement}
M. Gong thanks the support from the Max Planck Institute 
for extraterrestrial Physics and Princeton University.
X. Zheng is supported by the China Postdoctoral Science Foundation
(Grant No. 2017M610865).
D.N.C. Lin thanks IAS Princeton, IAS Tsinghua, IoA Cambridge, 
and ITC Harvard for support when this work was completed.
This work is partly supported by the National Key Basic Research and
Development Program of China (No. 2018YFA0404501 to SM) and
by the National Science Foundation of China 
(Grant No. 11333003, 11390372 and 11761131004 to S. Mao).
We also thank the anonymous referee for a helpful report.

\bibliographystyle{apj}
\bibliography{apj-jour,chondrule}
\end{document}